\documentclass[copyright,creativecommons]{eptcs}

\usepackage{graphicx}

\usepackage{pdfpages}

\usepackage{float}

\usepackage{amssymb}
\usepackage{amstext}
\usepackage{amsmath}

\providecommand{\event}{ThEdu'19 Post-Proceedings} 
\usepackage{underscore}           

\usepackage{hyperref}

\usepackage{graphicx}
\usepackage{proof}
\usepackage{minibox}

\usepackage{tikz}
\usetikzlibrary{tikzmark}
\usepackage{pstricks}

\usepackage{bussproofs}
\usepackage{ifthen}

\usepackage{listings}
\usepackage{quoting}

\newcommand{\citep}{\cite}

\newcommand{\tturnstile}{\vdash\kern-0.75em{\vdash}}
\newcommand{\eopl}{1em}
\newcommand{\eoplgrp}{2em}
\newcommand{\pgrptitle}{.75em}
\usepackage{booktabs}

\title{Teaching a Formalized Logical Calculus}

\def\titlerunning{Teaching a Formalized Logical Calculus}

\author{Asta Halkj{\ae}r From \email{} \and Alexander Birch Jensen \email{} \and Anders Schlichtkrull \email{} \and ~~~ J{\o}rgen Villadsen \footnote{Corresponding author: \texttt{jovi@dtu.dk}} \email{}
\institute{DTU Compute - Department of Applied Mathematics and Computer Science,\\[1ex]
Technical University of Denmark, Richard Petersens Plads, Building 324, DK-2800 Kongens Lyngby, Denmark}
}

\def\authorrunning{A. H. From, A. B. Jensen, A. Schlichtkrull \&\ J. Villadsen}

\begin{document}

\maketitle

\begin{abstract}
Classical first-order logic is in many ways central to work in mathematics, linguistics, computer science and artificial intelligence, so it is worthwhile to define it in full detail. We present soundness and completeness proofs of a sequent calculus for first-order logic, formalized in the interactive proof assistant Isabelle/HOL. Our formalization is based on work by Stefan Berghofer, which we have since updated to use Isabelle's declarative proof style Isar (Archive of Formal Proofs, Entry FOL-Fitting, August 2007 / July 2018). We represent variables with de Bruijn indices; this makes substitution under quantifiers less intuitive for a human reader.
However, the nature of natural numbers yields an elegant solution when compared to implementations of substitution using variables represented by strings. The sequent calculus considered has the special property of an always empty antecedent and a list of formulas in the succedent. We obtain the proofs of soundness and completeness for the sequent calculus as a derived result of the inverse duality of its tableau counterpart. We strive to not only present the results of the proofs of soundness and completeness, but also to provide a deep dive into a programming-like approach to the formalization of first-order logic syntax, semantics and the sequent calculus.
We use the formalization in a bachelor course on logic for computer science and discuss our experiences.
\end{abstract}

\section{Introduction}

Classical first-order logic is often used in mathematics, linguistics, philosophy and computer science. 
It is worthwhile to define it formally and recent advances in proof assistants like Isabelle/HOL have made it feasible \cite{IsaFoL}.

The sequent calculus considered has the special property of an always empty antecedent and a list of formulas in the succedent. We obtain the proofs of soundness and completeness for the sequent calculus as a derived result of the inverse duality of its tableau counterpart.

We strive to not only present the results of the proofs of soundness and completeness, but also to provide a deep dive into a programming-like approach to the formalization of first-order logic syntax, semantics and the sequent calculus. This is advantageous in a bachelor course, in particular when the students are familiar with functional programming.

We have taken the formalization of natural deduction by Berghofer \cite{berghofer} as the starting point, but we have updated the soundness and completeness proofs using Isabelle's declarative proofs style Isar \citep{wenzel99}.
The soundness and completeness proofs for tableaux and the sequent calculus are new for this paper.

We represent variables with de Bruijn indices; this makes substitution under quantifiers less intuitive for a human reader.
However, the nature of natural numbers yields an elegant solution when compared to implementations of substitution using variables represented by strings.

The development described in this paper is available online:

\begin{center}
{\small \url{https://bitbucket.org/isafol/isafol/src/master/FOL_Berghofer/}}
\end{center}

\begin{verbatim}
  4253 lines FOL_Berghofer.thy
   867 lines FOL_Tableau.thy     A contribution of this paper
   217 lines FOL_Sequent.thy     A contribution of this paper
   132 lines FOL_Appendix.thy    A contribution of this paper
\end{verbatim}

These numbers include blank lines and a few comments.
All in all it takes around 5 seconds in real time to verify on a fairly standard computer.
The entire formalization is based on the standard theory \texttt{Main} (the standard library of Isabelle/HOL which comes with many useful functions and facts about e.g.\ natural numbers and lists).

We have recently named our system based on the formalization in Isabelle/HOL:

\noindent\includegraphics[trim=-60mm 170mm 60mm 12mm,clip,width=
\textwidth,height=\textheight,keepaspectratio,page=1]{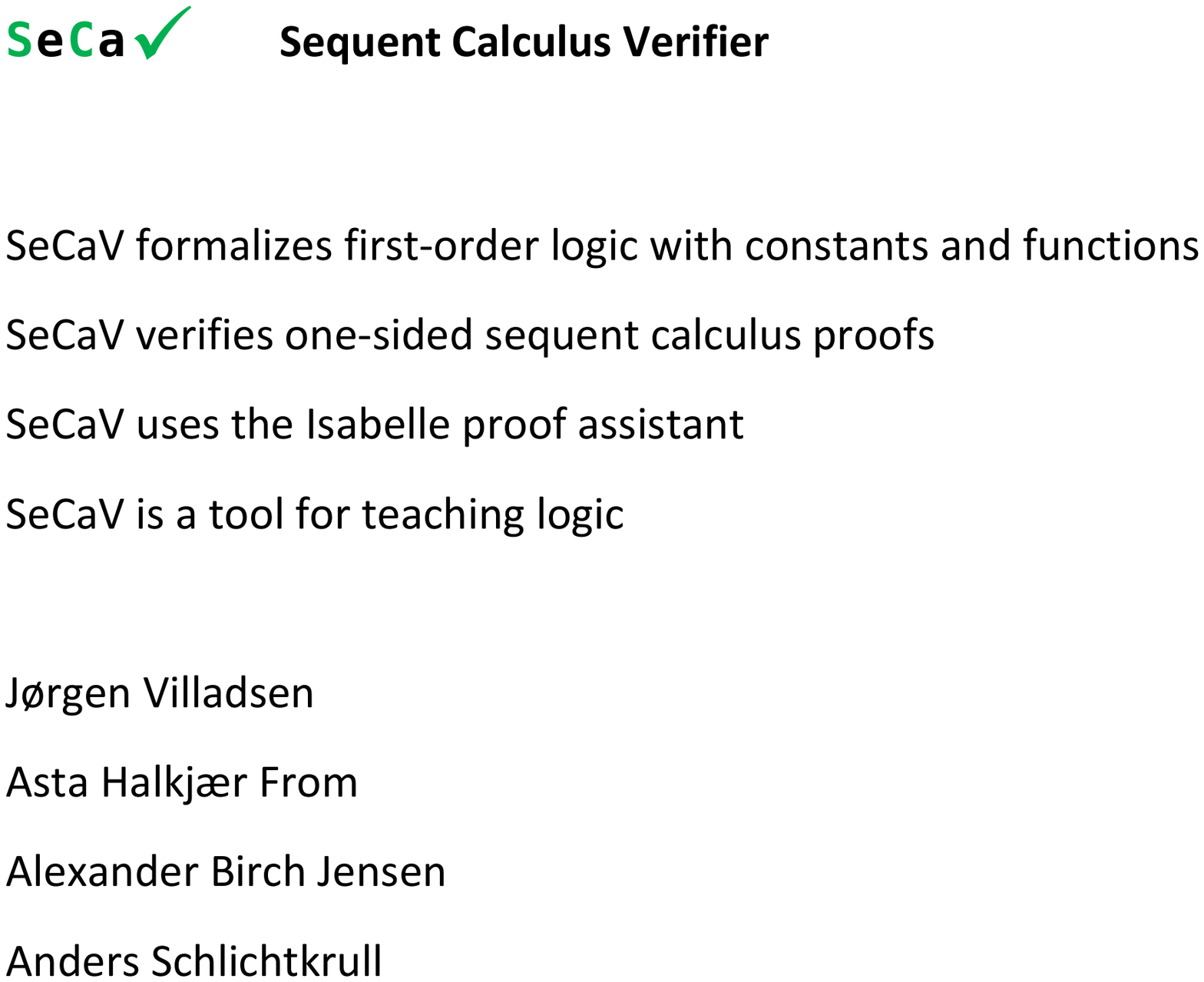}

For our bachelor course we focus on the fragment (499 lines) of the formalization available here:
\begin{center}
{\small \url{https://bitbucket.org/isafol/isafol/src/master/Sequent_Calculus/}}
\end{center}
Here only the soundness proof is included but in addition a small and a large proof in the sequent calculus is formalized.

The structure of the paper is as follows.
Section 2 explains the formalization of the syntax and the semantics of classical first-order logic.
Section 3 describes the sequent calculus.
Section 4 outlines the formalized soundness and completeness proofs.
Section 5 considers teaching the sequent calculus. 
Section 6 is a discussion of related work.
Finally, section 7 is the conclusion.

\section{Formalization of Syntax and Semantics}

We provide in this section definitions for the syntax and semantics of first-order classical logic without equality. This part of the formalization is based on the work by Berghofer \cite{berghofer}. 

Throughout the presentation of our formalization, we will state and explain the types of definitions and functions. In particular, the use of type variables requires further explanation. In Isabelle a type variable states an arbitrary type. We will consistently use the type variables \texttt{'a}, \texttt{'b} and \texttt{'c} for our model. The type \texttt{'a} specifies the type of function identifiers, e.g.\ strings or natural numbers. The type \texttt{'b} similarly specifies the type of predicate identifiers. Finally, the type \texttt{'c} specifies the type of elements in the universe.

\subsection{Syntax of first-order logic formulas}
This section provides the necessary definitions to construct well-formed formulas of classical first-order logic without equality. We start by defining the syntax for terms of first-order logic:
\begin{lstlisting}
term ::=
\end{lstlisting}
\vspace{-1ex}
\begin{lstlisting}
    Var nat
    Fun 'a [term, ..., term]
\end{lstlisting}
The terms are variables and functions as usual.
The type \texttt{nat} is for natural numbers as we use de Bruijn indices and thus variables appear as natural numbers.
Constants are represented as functions with no arguments.

We now turn to define the syntax for formulas of first-order logic:
\begin{lstlisting}
form ::= 
\end{lstlisting}
\vspace{-1ex}
\begin{lstlisting}
  $\bot$ 
  $\top$ 
  Pre 'b ['a term, ..., 'a term] 
  Con form form 
  Dis form form 
  Imp form form 
  Neg form
  Uni form
  Exi form
\end{lstlisting}
Formulas have the usual first-order logic connectives and also include $\bot$, $\top$ and $\neg$. The constructor \texttt{Pre} is for predicates, \texttt{Con} is for $\land$, \texttt{Dis} is for $\lor$, \texttt{Imp} is for $\longrightarrow$, \texttt{Neg} is for $\neg$, \texttt{Uni} is for the universal quantifier $\forall$, and \texttt{Exi} is for the existential quantifier $\exists$. 

It must be emphasized that all functions and predicates have arbitrarily many arguments.

The use of de Bruijn indices in formulas is most easily explained by an example:

\[
g(\tikzmark{w}z) \longrightarrow (\forall \tikzmark{x1}x.~ p(\tikzmark{x2}x) \longrightarrow (\exists \tikzmark{y1}y.~ q(\tikzmark{y2}y) \longrightarrow (p(\tikzmark{x3}x) \lor q(\tikzmark{w2}z))))
\]

\begin{tikzpicture}[remember picture, overlay]
\draw[<-] 
  ([shift={(2pt,-3pt)}]pic cs:w) |- ([shift={(-7pt,-10pt)}]pic cs:w) 
  node[anchor=east] {$\scriptstyle \text{free variable}$}; 
\draw[<-] 
  ([shift={(2pt,-3pt)}]pic cs:w2) |- ([shift={(14pt,-10pt)}]pic cs:w2) 
  node[anchor=west] {$\scriptstyle \text{free variable}$}; 
\draw[<->] 
  ([shift={(3pt,-3pt)}]pic cs:x1) |- ([shift={(3pt,-18pt)}]pic cs:x2) 
  |- ([shift={(3pt,-3pt)}]pic cs:x2)
  node[anchor=west] {}; 
\draw[->] 
  ([shift={(3pt,-3pt)}]pic cs:x2) |- ([shift={(3pt,-18pt)}]pic cs:x3) 
  |- ([shift={(3pt,-3pt)}]pic cs:x3)
  node[anchor=west] {}; 
\draw[<->] 
  ([shift={(4pt,-5pt)}]pic cs:y1) |- ([shift={(3pt,-12pt)}]pic cs:y2) 
  |- ([shift={(3pt,-5pt)}]pic cs:y2)
  node[anchor=west] {};
\end{tikzpicture}\vspace{1em}

The arrows indicate the references of variables with regard to quantifiers and free variables. Consider below the same formula using de Bruijn indices instead: 

\[
g(\tikzmark{dbw}0) \longrightarrow ( \tikzmark{dbx1}\forall.~ p(\tikzmark{dbx2}0) \longrightarrow ( \tikzmark{dby1}\exists.~ q(\tikzmark{dby2}0) \longrightarrow (p(\tikzmark{dbx3}1) \lor q(\tikzmark{dbz}2))))
\]

\begin{tikzpicture}[remember picture, overlay]
\draw[<-] 
  ([shift={(2pt,-3pt)}]pic cs:dbw) |- ([shift={(-7pt,-10pt)}]pic cs:dbw) 
  node[anchor=east] {$\scriptstyle \text{free variable}$}; 
\draw[<-] 
  ([shift={(3pt,-3pt)}]pic cs:dbz) |- ([shift={(14pt,-10pt)}]pic cs:dbz) 
  node[anchor=west] {$\scriptstyle \text{free variable}$}; 
\draw[<->] 
  ([shift={(3pt,-3pt)}]pic cs:dbx1) |- ([shift={(3pt,-18pt)}]pic cs:dbx2) 
  |- ([shift={(3pt,-3pt)}]pic cs:dbx2)
  node[anchor=west] {};
\draw[->] 
  ([shift={(3pt,-3pt)}]pic cs:dbx2) |- ([shift={(3pt,-18pt)}]pic cs:dbx3) 
  |- ([shift={(3pt,-3pt)}]pic cs:dbx3)
  node[anchor=west] {};
\draw[<->] 
  ([shift={(4pt,-5pt)}]pic cs:dby1) |- ([shift={(3pt,-12pt)}]pic cs:dby2) 
  |- ([shift={(3pt,-5pt)}]pic cs:dby2)
  node[anchor=west] {};
\end{tikzpicture}\vspace{1em}

Syntactic representations of quantifiers such as $\exists x$ are replaced by $\exists$ as the variable referencing is implicit. A variable $0$ is bound by the innermost quantifier, a variable $1$ inside two quantifiers is bound by the outermost quantifier, and so on. A variable references a free variable when its index exceeds or equals the number of quantifiers it is bound by. Outside the scope of any quantifier, $0$ references a free variable, $1$ another free variable, and so on.

We define the syntax of Herbrand terms that are closed by construction (ground terms).
\begin{lstlisting}
hterm ::= HFun 'a [hterm, ..., hterm]
\end{lstlisting}
Defining the Herbrand terms as a separate type plays well with our use of a type variable to represent the universe, which for the completeness theorem consists of Herbrand terms. Herbrand terms are distinguished from regular terms by having no variables.
Alternatively we could introduce and reason about an explicit sub-type of the regular terms.

\subsection{Semantics of first-order logic}

We present here a formalization of the semantics of first-order classical logic. The semantics describe the truth evaluation of formulas within a given model i.e. for a given universe, environment and interpretation of function and predicate identifiers.

In the context of programming in Isabelle, it is particularly important to understand the components of a model. The environment $e$ maps variables to elements of the universe. The type of elements in the universe is arbitrary and the variables are natural numbers.

The interpretation $f$ maps function identifiers to functions on the universe and $g$ maps predicate identifiers to predicates. Combined they form our model. Given an instance of such a model we can evaluate formulas.

The semantics of first-order formulas is defined below as the function \texttt{semantics} along with functions \texttt{semantics\_term} and \texttt{semantics\_list} for the semantics of terms and lists of terms, respectively:
\begin{lstlisting}
semantics_term :: (nat $\Rightarrow$ 'c) $\Rightarrow$ ('a $\Rightarrow$ 'c list $\Rightarrow$ 'c) $\Rightarrow$ term $\Rightarrow$ 'c
\end{lstlisting}
\vspace{-1ex}
\begin{lstlisting}
  semantics_term e f (Var n) = e n
  semantics_term e f (Fun i l) = f i (semantics_list e f l)

semantics_list :: (nat $\Rightarrow$ 'c) $\Rightarrow$ ('a $\Rightarrow$ 'c list $\Rightarrow$ 'c) $\Rightarrow$ 
    term list $\Rightarrow$ 'c list
\end{lstlisting}
\vspace{-1ex}
\begin{lstlisting}
  semantics_list e f [] = []
  semantics_list e f (t # l) = 
    semantics_term e f t # semantics_list e f l
\end{lstlisting}

Let us first inspect the types of our environment \texttt{e} and interpretation \texttt{f}. The environment has the type \texttt {(nat $\Rightarrow$ 'c)} as we map variables (natural numbers) to elements of the universe, i.e. elements of type \texttt{'c}. The interpretation of function identifiers \texttt{f} has the type \texttt{('a $\Rightarrow$ 'c list $\Rightarrow$ 'c)}. We map function identifiers of type \texttt{'a} along with its arguments to elements of the universe. Recall that arguments to functions are terms which themselves are mapped to elements of the universe by interpretation \texttt{f}. Therefore, we need not only the function identifier, but also the interpretation of its arguments.
For lists of terms \texttt{semantics\_list}, we simply map each element to its semantic value recursively. \begin{lstlisting}
semantics :: (nat $\Rightarrow$ 'c) $\Rightarrow$ ('a $\Rightarrow$ 'c list $\Rightarrow$ 'c) $\Rightarrow$ 
    ('b $\Rightarrow$ 'c list $\Rightarrow$ bool) $\Rightarrow$ form $\Rightarrow$ bool
\end{lstlisting}
\vspace{-1ex}
\begin{lstlisting}
  semantics e f g $\bot$ = False
  semantics e f g $\top$ = True
  semantics e f g (Pre i l) = g i (semantics_list e f l)
  semantics e f g (Con p q) = (semantics e f g p $\land$ semantics e f g q)
  semantics e f g (Dis p q) = (semantics e f g p $\lor$ semantics e f g q)
  semantics e f g (Imp p q) = (semantics e f g p $\longrightarrow$ semantics e f g q)
  semantics e f g (Neg p) = ($\neg$ semantics e f g p)
  semantics e f g (Uni p) = ($\forall$z. semantics (shift e 0 z) f g p)
  semantics e f g (Exi p) = ($\exists$z. semantics (shift e 0 z) f g p)
\end{lstlisting}

For the semantics of formulas we further have as an argument the predicate interpretation \texttt{g} of type \texttt{('b $\Rightarrow$ 'c list $\Rightarrow$ bool)}. Similarly to the function interpretation we map predicate identifiers of type \texttt{'b} along with its interpreted term arguments of type \texttt{'c list} to Boolean values.
The cases for most of the logical connectives are rather trivial. The explicit use of logical connectives on the right hand side of our function definition may at first seem rather confusing. However, these connectives should be understood within the context of Isabelle programming, namely as built-in functions that return the Boolean value of the operator applied to its arguments.

The cases of the existential and universal quantifiers are more challenging. Consider the quantifiers on the right-hand side as programmable functions that work on a set of possible values of \texttt{z}. By inspection in Isabelle, and later through understanding the function \texttt{shift}, we realize that the type of \texttt{z} is \texttt{'c}. This means that the existential quantifier checks if there is an element in the universe that makes the formula \texttt{p} true. Similarly, the universal quantifier checks if all elements make \texttt{p} true. Exactly how we put in $z$ at the places of the quantified variables is handled by \texttt{shift}, which is defined below:
\begin{lstlisting}
shift :: (nat $\Rightarrow$ 'c) $\Rightarrow$ nat $\Rightarrow$ 'c $\Rightarrow$ (nat $\Rightarrow$ 'c)
\end{lstlisting}
\vspace{-1ex}
\begin{lstlisting}
  shift e v z $\equiv$ 
    ($\lambda$n. if n < v then e n else if n = v then z else e (n - 1))
\end{lstlisting}

We can consider \texttt{shift} a function that takes as arguments an existing environment \texttt{e} of type \texttt{(nat $\Rightarrow$ 'c)}, a variable \texttt{v} of type \texttt{nat} and an element of the universe \texttt{z} of type \texttt{'c}, and the result is an updated environment. In the semantics, \texttt{shift} is always used with \texttt{v} having the value 0. This essentially boils down the line we have to understand to:
\begin{lstlisting}
  shift e 0 z $\equiv$ 
    ($\lambda$n. if n < 0 then e n else if n = 0 then z else e (n - 1))
\end{lstlisting}
Since \texttt{n} is a natural number it further reduces to:
\begin{lstlisting}
  shift e 0 z $\equiv$ 
    ($\lambda$n. if n = 0 then z else e (n - 1))
\end{lstlisting}

Recall how de Bruijn indices are used under quantifiers. When checking if the quantifier holds for a given \texttt{z}, we swap in \texttt{z} for all variables with index 0. In the process of evaluating we essentially eliminate the quantifier, as we now just consider one specific element \texttt{z} of the universe, and hence indices above 0 are subtracted by 1 to adjust variables that were previously within the scope of a quantifier.

The complicated definition of shift is preferred instead of the simpler one since it matches the pattern used in the definition of substitution to be explained later.

\section{The Proof System}

In this section we go over an informal definition of the sequent calculus. We present the proof system in its entirety in a format one would expect from a textbook. Simultaneously, we present in slightly abbreviated Isabelle syntax the rules of the proof system, and we explain the rules in detail. 

For general sequent calculi the most common way of writing a proof state is $$\Gamma \vdash \Delta$$ where $\Gamma = \left\{\gamma_1, \ldots, \gamma_n \right\}$ and $\Delta = \left\{\delta_1, \ldots, \delta_n \right\}$ are sets of formulas. The operator $\vdash$ is to be interpreted as an implication with the conjunction of formulas in $\Gamma$ on the left-hand side and the disjunction of the formulas in $\Delta$ on the right-hand side: 
$$ \left( \gamma_1 \land \ldots \land \gamma_n \right) \longrightarrow \left( \delta_1 \lor \ldots \lor \delta_n \right) $$
In our case the sets of formulas $\Gamma$ and $\Delta$ are lists of formulas and $\Gamma$ is always empty (this restriction makes the system simpler). So instead of the symbol $\vdash$ we use the symbol \lstinline'$\tturnstile$' and lists of formulas are preferred rather than sets or multisets since lists are well-known from functional programming and always finite.

For certain rules, especially those regarding quantifiers, side conditions apply. In particular, we have conditions regarding substitution under quantifiers and newness of constants and functions. Furthermore, we require functions for list membership and extension. We go through each of the required auxiliary functions and discuss their details.

\subsection{Newness of Constants and Functions}
Recall that constants are nothing but functions without arguments.
For the introduction of a universal quantifier $\forall$ and its negated counterpart $\neg \exists$, we have requirements of newness for the instantiated terms. For a function identifier to be new, it must not occur in any of the formulas in the succedent. Below we define functions for checking newness of a constant or a function with respect to a term and a list of terms, respectively:

\begin{lstlisting}
new_term :: 'a $\Rightarrow$ term $\Rightarrow$ bool 
\end{lstlisting}
\vspace{-1ex}
\begin{lstlisting}
  new_term c (Var n) = True
  new_term c (Fun i l) = (if i = c then False else new_list c l)
\end{lstlisting}
\begin{lstlisting}
new_list :: 'a $\Rightarrow$ term list $\Rightarrow$ bool
\end{lstlisting}
\vspace{-1ex}
\begin{lstlisting}
  new_list c [] = True
  new_list c (t # l) = (if new_term c t then new_list c l else False)
\end{lstlisting}

Given a function identifier \texttt{c} of type \texttt{'a} and a term \texttt{t}, we can trivially determine if \texttt{c} is new by checking if \texttt{t} is in fact a function with identifier \texttt{i} and if \texttt{i = c}. For checking a list of terms, we recursively traverse the list and check if newness holds for all terms in the list.

What we call newness is often called freshness but in general we prefer the shorter word and always make everything very explicit.

Note that we here and elsewhere prefer to use if-then-else constructs instead of conjunctions in order to avoid any confusions with the logical connectives of first-order logic.

We now define newness of constants and functions in formulas:
\begin{lstlisting}
new :: 'a $\Rightarrow$ form $\Rightarrow$ bool
\end{lstlisting}
\vspace{-1ex}
\begin{lstlisting}
  new c $\bot$ = True
  new c $\top$ = True
  new c (Pre i l) = new_list c l
  new c (Con p q) = (if new c p then new c q else False)
  new c (Dis p q) = (if new c p then new c q else False)
  new c (Imp p q) = (if new c p then new c q else False)
  new c (Neg p) = new c p
  new c (Uni p) = new c p
  new c (Exi p) = new c p
\end{lstlisting}

We also define newness of constants and functions in lists of formulas:
\begin{lstlisting}
news :: 'a $\Rightarrow$ form list $\Rightarrow$ bool
\end{lstlisting}
\vspace{-1ex}
\begin{lstlisting}
  news c [] = True
  news c (p # x) = (if new c p then news c x else False)
\end{lstlisting}
The types of \texttt{new} and \texttt{news} are almost identical to \texttt{new\_term} and \texttt{new\_list}, barring the fact that they take a formula as input instead of a term. While most cases require no further explanation, observe that for the binary connectives the term has to be new in both the left-hand and right-hand side. A predicate is clearly the only case where terms appear directly, and we here use the previously defined function  \texttt{new\_list}. 
Again, we traverse the list recursively to check if newness holds for all formulas in the list.

\subsection{Substitution of Variables}
For any rule that introduces a quantifier, correct substitution of quantified variables is an integral part of the proof step. Before we define substitution for formulas, we first define a function that increments all variables of a term by 1. 
\begin{lstlisting}
inc_term :: term $\Rightarrow$ term
\end{lstlisting}
\vspace{-1ex}
\begin{lstlisting}
  inc_term (Var n) = Var (n + 1)
  inc_term (Fun i l) = Fun i (inc_list l)
\end{lstlisting}
Given a term as input, the result is a new term with any de Bruijn indices incremented by one. 
We also define the function for a term list:
\begin{lstlisting}
inc_list :: term list $\Rightarrow$ term list
\end{lstlisting}
\vspace{-1ex}
\begin{lstlisting}
  inc_list [] = []
  inc_list (t # l) = inc_term t # inc_list l
\end{lstlisting}
Furthermore, we define substitution in a term as its own function:
\begin{lstlisting}
sub_term :: nat $\Rightarrow$ term $\Rightarrow$ term $\Rightarrow$ term
\end{lstlisting}
\vspace{-1ex}
\begin{lstlisting}
  sub_term v s (Var n) = 
        (if n < v then Var n else if n = v then s else Var (n - 1))
  sub_term v s (Fun i l) = Fun i (sub_list v s l)
\end{lstlisting}
As indicated by the type, we have as a natural number \texttt{v}, the term we substitute as \texttt{s}, and the term in which the variable with de Bruijn index \texttt{v} is to be substituted. Due to the recursive definition, only the case where this term is a variable gives rise to a substitution.
For terms alone we ignore any possible quantifiers. As such, a substitution inside a term can be seen as replacing all free variables with de Bruijn index \texttt{v}. Consequently, we decrement all indices greater than \texttt{v} while leaving indices below \texttt{v} untouched.

We perform substitution on a list of terms by traversing the list recursively:

\begin{lstlisting}
sub_list :: nat $\Rightarrow$ term $\Rightarrow$ term list $\Rightarrow$ term list
\end{lstlisting}
\vspace{-1ex}
\begin{lstlisting}
  sub_list v s [] = []
  sub_list v s (t # l) = sub_term v s t # sub_list v s l
\end{lstlisting}
We can now define substitution in a formula as a function:

\begin{lstlisting}
sub :: nat $\Rightarrow$ term $\Rightarrow$ form $\Rightarrow$ form
\end{lstlisting}
\vspace{-1ex}
\begin{lstlisting}
  sub v s $\bot$ = $\bot$
  sub v s $\top$ = $\top$
  sub v s (Pre i l) = Pre i (sub_list v s l)
  sub v s (Con p q) = Con (sub v s p) (sub v s q)
  sub v s (Dis p q) = Dis (sub v s p) (sub v s q)
  sub v s (Imp p q) = Imp (sub v s p) (sub v s q)
  sub v s (Neg p) = Neg (sub v s p)
  sub v s (Uni p) = Uni (sub (v + 1) (inc_term s) p)
  sub v s (Exi p) = Exi (sub (v + 1) (inc_term s) p)
\end{lstlisting}
As in \texttt{sub\_term} we have as input a natural number \texttt{v}, a term \texttt{s} and a formula in which the substitution is to be performed. The cases for $\bot$ and $\top$ are trivial, and the cases for the logical connectives \texttt{Con}, \texttt{Dis}, \texttt{Imp} and \texttt{Neg} simply call \texttt{sub} recursively on their arguments. Substitution for terms is used on predicate arguments in the case for \texttt{Pre}. 
The two cases for the quantifiers require closer examination. We observe that they are almost identical. We reconstruct the formula using \texttt{sub} recursively in the constructor argument. In comparison to the other cases, we have \texttt{v + 1} instead of \texttt{v} and \texttt{inc\_term s} instead of \texttt{s}. This is due to the fact that all de Bruijn indices are incremented when inside quantifiers. As previously described, a free variable \texttt{Var 0} will be written \texttt{Var 1} when inside a single quantifier. Similarly, we increment variables in the term that is to be substituted in when encountering a quantifier.

\subsection{Extending a List of Formulas}
As we will uncover later, a rule in the sequent calculus allows for an extension of the list of formulas in the succedent by any number of formulas. This is basically exchange, contraction, and weakening as standard in sequent calculus. We program this as a function \texttt{ext} using the following \texttt{member} function:
\begin{lstlisting}
member :: form $\Rightarrow$ form list $\Rightarrow$ bool
\end{lstlisting}
\vspace{-1ex}
\begin{lstlisting}
  member p [] = False
  member p (q # x) = (if p = q then True else member p x)
\end{lstlisting}
Hence \texttt{member p x} returns \texttt{True} if the formula \texttt{p} occurs in \texttt{x}.

\begin{lstlisting}
ext :: form list $\Rightarrow$ form list $\Rightarrow$ bool
\end{lstlisting}
\vspace{-1ex}
\begin{lstlisting}
  ext y [] = True
  ext y (p # x) = (if member p y then ext y x else False)
\end{lstlisting}
Hence \texttt{ext y x} returns \texttt{True} if every formula in \texttt{x} is a member of \texttt{y}. 
The extension function \texttt{ext} is simpler than a permutation function (which would be sufficient).

\subsection{Inductive Definition of Sequent Calculus}

We present here the inductive definition of the sequent calculus proof system. In Figure \ref{fig:proofsystem-inf} all rules are presented informally in a format well-known to readers of modern logic books. 
\begin{figure*}
\begin{center}
\begin{minipage}{\textwidth}
\centering \textbf{Leaf rules}
\end{minipage}\vspace{\pgrptitle}
\begin{minipage}{\textwidth}
\begin{prooftree}
\AxiomC{}
\UnaryInfC{$\tturnstile P(v_{1}, \ldots ,v_{k}),\neg P(v_{1}, \ldots ,v_{k}),\Delta$}
\end{prooftree}
\end{minipage}\vspace{\eopl}
\begin{minipage}{.175\textwidth}
\begin{prooftree}
\AxiomC{}
\UnaryInfC{$\tturnstile \neg \bot, \Delta$}
\end{prooftree}
\end{minipage}
\begin{minipage}{.175\textwidth}
\begin{prooftree}
\AxiomC{}
\UnaryInfC{$\tturnstile \top,\Delta$}
\end{prooftree}
\end{minipage}\vspace{\eoplgrp}
\begin{minipage}{\textwidth}
\centering \textbf{$\alpha$-rules}
\end{minipage}\vspace{\pgrptitle}
\begin{minipage}{.33\textwidth}
\begin{prooftree}
\AxiomC{$\tturnstile p, \Delta$}
\UnaryInfC{$\tturnstile \neg\neg p, \Delta$}
\end{prooftree}
\end{minipage}
\begin{minipage}{.33\textwidth}
\begin{prooftree}
\AxiomC{$\tturnstile \neg p, q, \Delta$}
\UnaryInfC{$\tturnstile p \longrightarrow q, \Delta $}
\end{prooftree}
\end{minipage}\vspace{\eopl}
\begin{minipage}{.33\textwidth}
\begin{prooftree}
\AxiomC{$\tturnstile \neg p, \neg q, \Delta$}
\UnaryInfC{$\tturnstile \neg (p \land q), \Delta$}
\end{prooftree}
\end{minipage}
\begin{minipage}{.33\textwidth}
\begin{prooftree}
\AxiomC{$\tturnstile p, q,\Delta$}
\UnaryInfC{$\tturnstile p \lor q, \Delta$}
\end{prooftree}
\end{minipage}\vspace{\eoplgrp}
\begin{minipage}{\textwidth}
\centering \textbf{$\beta$-rules}
\end{minipage}\vspace{\pgrptitle}
\begin{minipage}{.33\textwidth}
\begin{prooftree}
\AxiomC{$\tturnstile p, \Delta$}
\AxiomC{$\tturnstile q, \Delta$}
\BinaryInfC{$\tturnstile p \land q ,\Delta$}
\end{prooftree}
\end{minipage}
\begin{minipage}{.33\textwidth}
\begin{prooftree}
\AxiomC{$\tturnstile \neg p , \Delta$}
\AxiomC{$\tturnstile \neg q , \Delta$}
\BinaryInfC{$\tturnstile \neg (p \lor q),\Delta$}
\end{prooftree}
\end{minipage}\vspace{\eopl}
\begin{minipage}{.33\textwidth}
\begin{prooftree}
\AxiomC{$\tturnstile p, \Delta$}
\AxiomC{$\tturnstile \neg q, \Delta$}
\BinaryInfC{$\tturnstile \neg (p \longrightarrow q) ,\Delta$}
\end{prooftree}
\end{minipage}\vspace{\eoplgrp}
\begin{minipage}{\textwidth}
\centering \textbf{$\Delta$-rules}
\end{minipage}\vspace{\pgrptitle}
\begin{minipage}{.5\textwidth}
\begin{prooftree}
\AxiomC{$\tturnstile p[t/0], \Delta$}
\UnaryInfC{$\tturnstile (\exists . p),\Delta$}
\end{prooftree}
\end{minipage}
\begin{minipage}{.42\textwidth}
$p[t/0]$ is the formula $p$ with the variable $0$ substituted by the term $t$.
\end{minipage}\vspace{\eopl}
\begin{minipage}{.5\textwidth}
\begin{prooftree}
\AxiomC{$\tturnstile \neg p[t/0], \Delta$}
\UnaryInfC{$\tturnstile \neg (\forall . p),\Delta$}
\end{prooftree}
\end{minipage}
\begin{minipage}{.42\textwidth}
$p[t/0]$ is the formula $p$ with the variable $0$ substituted by the term $t$.
\end{minipage}\vspace{\eoplgrp}
\begin{minipage}{\textwidth}
\centering \textbf{$\delta$-rules}
\end{minipage}\vspace{\pgrptitle}
\begin{minipage}{.5\textwidth}
\begin{prooftree}
\AxiomC{$\tturnstile p[t/0], \Delta$}
\UnaryInfC{$\tturnstile (\forall . p),\Delta$}
\end{prooftree}
\end{minipage}
\begin{minipage}{.42\textwidth}
$p[t/0]$ is the formula $p$ with the variable $0$ substituted by the term $t$ (a fresh constant).
\end{minipage}\vspace{\eopl}
\begin{minipage}{.5\textwidth}
\begin{prooftree}
\AxiomC{$\tturnstile \neg p[t/0], \Delta$}
\UnaryInfC{$\tturnstile \neg (\exists. p),\Delta$}
\end{prooftree}
\end{minipage}
\begin{minipage}{.42\textwidth}
$p[t/0]$ is the formula $p$ with the variable $0$ substituted by the term $t$ (a fresh constant).
\end{minipage}\vspace{\eoplgrp}
\begin{minipage}{\textwidth}
\centering \textbf{Extension rule}
\end{minipage}\vspace{\pgrptitle}
\begin{minipage}{.5\textwidth}
\begin{prooftree}
\AxiomC{$\tturnstile x$}
\UnaryInfC{$\tturnstile y$}
\end{prooftree}
\end{minipage}
\begin{minipage}{.42\textwidth}
Every element in the list of formulas $x$ must be a member of $y$.
\end{minipage}
\end{center}
\caption{Proof System}
\label{fig:proofsystem-inf}
\end{figure*}
We have already covered all the auxiliary functions needed for programming the side conditions of difficult rules. The inductive definition for the entire sequent calculus proof system is given below:

\begin{lstlisting}
sequent_calculus ($\tturnstile$) :: ('a, 'b) form list $\Rightarrow$ bool
\end{lstlisting}
\vspace{-1ex}
\begin{lstlisting}
  $\tturnstile$ Pre i l # Neg (Pre i l) # x 
  $\tturnstile$ Neg $\bot$ # x 
  $\tturnstile$ $\top$ # x 
  $\tturnstile$ p # x $\Longrightarrow$ $\tturnstile$ Neg (Neg p) # x 
  $\tturnstile$ Neg p # Neg q # x $\Longrightarrow$ $\tturnstile$ Neg (Con p q) # x 
  $\tturnstile$ p # q # x $\Longrightarrow$ $\tturnstile$ Dis p q # x 
  $\tturnstile$ Neg p # q # x $\Longrightarrow$ $\tturnstile$ Imp p q # x 
  $\tturnstile$ p # x $\Longrightarrow$ $\tturnstile$ q # x $\Longrightarrow$ $\tturnstile$ Con p q # x 
  $\tturnstile$ Neg p # x $\Longrightarrow$ $\tturnstile$ Neg q # x $\Longrightarrow$ $\tturnstile$ Neg (Dis p q) # x 
  $\tturnstile$ p # x $\Longrightarrow$ $\tturnstile$ Neg q # x $\Longrightarrow$ $\tturnstile$ Neg (Imp p q) # x 
  $\tturnstile$ sub 0 t p # x $\Longrightarrow$ $\tturnstile$ Exi p # x 
  $\tturnstile$ Neg (sub 0 t p) # x $\Longrightarrow$ $\tturnstile$ Neg (Uni p) # x 
  $\tturnstile$ sub 0 (Fun i []) p # x $\Longrightarrow$ news i (p # x) $\Longrightarrow$ $\tturnstile$ Uni p # x 
  $\tturnstile$ Neg (sub 0 (Fun i []) p) # x $\Longrightarrow$ news i (p # x) $\Longrightarrow$ $\tturnstile$ Neg (Exi p) # x 
  $\tturnstile$ x $\Longrightarrow$ ext y x $\Longrightarrow$ $\tturnstile$ y
\end{lstlisting}

We explain here only the most difficult rules. Note that the operator $\Longrightarrow$ is a meta-implication operator in Isabelle. The meta-implication operates on a higher-level than the usual logical operator $\longrightarrow$. The rules are inductively defined by use of this operator and appeal to a top-down construction of proofs, but it can be useful to consider proof construction in the reverse direction as well. Multiple uses of $\Longrightarrow$ indicate that the conclusion follows from multiple assumptions. When constructing the proof bottom-up, leaf rules mark the end of a branch in the proof tree. We have three leaf rules. Two of them are for $\top$ and $\neg \bot$. If either $\top$ or $\neg \bot$ is the first formula in the succedent, clearly its disjunction is always true. The third and possibly most interesting leaf rule is well-known from Gentzen-style systems:
\begin{lstlisting}
$\tturnstile$ Pre i l # Neg (Pre i l) # x
\end{lstlisting}
The rule states that we can end a branch in the proof tree if a predicate appears as the first formula of the succedent and if the second formula is the negation of the same predicate (with the same arguments). In other similar systems there are no requirements of the order of a predicate and its negation. As we shall see later, this extra condition is circumvented by application of the extension rule. As we are under the assumptions of classical logic, it is trivial to argue that either a predicate or its negation must be true for the same arguments.

The next rule highlighted is for introduction of the existential quantifier $\exists$:
\begin{lstlisting}
$\tturnstile$ sub 0 t p # x $\Longrightarrow$ $\tturnstile$ Exi p # x
\end{lstlisting}
We may existentially quantify the formula over all occurrences of a term \texttt{t}.

The following rule is for introduction of a universal quantifier $\forall$:
\begin{lstlisting}
$\tturnstile$ sub 0 (Fun i []) p # x $\Longrightarrow$ news i (p # x) $\Longrightarrow$ $\tturnstile$ Uni p # x
\end{lstlisting}
Here it is insufficient to require that the formula is provable for some term \texttt{t}. We additionally require that the term must be an arbitrarily chosen constant \texttt{i} not previously used in the proof.

Finally we highlight the extension rule:
\begin{lstlisting}
$\tturnstile$ x $\Longrightarrow$ ext y x $\Longrightarrow$ $\tturnstile$ y
\end{lstlisting}
Recall that the function \texttt{ext y x} is only true if all formulas in \texttt{x} occur in \texttt{y}. However, there may be additional formulas in \texttt{y} and the ordering can be different. The main purpose of the rule is to rearrange the formulas in the succedent, seeing that every other rule merely considers the first formula in the succedent. 

We invite the reader to study the remaining rules of the sequent calculus.

\section{Proofs of Soundness and Completeness}
Previous sections have covered the formalization of syntax and semantics for classical first-order logic as well as the sequent calculus proof system in the Isabelle/HOL proof assistant. What remains is to prove key properties of the proof system. We present the most important theorems, namely soundness and completeness, and discuss their significance.
The proofs are rather extensive, but we discuss them briefly.

Soundness and completeness theorems revolve around the relation between the formulas \texttt{p} that are valid, written \texttt{semantics e f g p} where \texttt{e}, \texttt{f} and \texttt{g} are universally quantified, and those that can be proved in the sequent calculus, written \texttt{$\tturnstile$ p}.

For the completeness theorem we consider first a restricted form of validity:
\begin{lstlisting}
 ($\gg$ p :: (nat, nat) form) $\equiv$ 
    $\forall$(e :: _ $\Rightarrow$ nat hterm) f g. semantics e f g p
\end{lstlisting}
The universe consists of Herbrand terms using natural numbers as function identifiers as indicated by the type of the environment \texttt{nat $\Rightarrow$ nat hterm}. Furthermore, we restrict the regular function identifiers and predicate identifiers to natural numbers as well, as indicated by the type of formulas \texttt{(nat, nat) form}.

This restricted form of validity is used for the completeness theorem:
\begin{lstlisting}
$\gg$ p $\Longrightarrow$ $\tturnstile$ [p]
\end{lstlisting}
The theorem states that if a formula is valid, in the restricted form, then it can also be proved in the sequent calculus. The proof can only be completed in the current setup if we use a restricted validity term. From a theoretical point of view, however, restricting the universe, the function identifiers and the predicate identifiers to specific types does not weaken the completeness result. In fact, showing completeness under these assumptions is theoretically more challenging. This is best explained by an argument: Consider the amount of formulas that are valid in all universes. Now also consider the amount of formulas that are valid in a specific universe. Clearly, every formula that is valid in all universes is also valid in a specific universe. However, the specific universe may have additional valid formulas. Consequently, if we show that all valid formulas of a specific universe can be proved, we have already proved all the formulas that are valid in all universes. 

The soundness theorem is proved for any model:
\begin{lstlisting}
$\tturnstile$ [q] $\Longrightarrow$ semantics e f g q
\end{lstlisting}
The theorem states that the sequent calculus only proves valid formulas. Combining the soundness and completeness result we can conclude that the implication holds in both directions:
\begin{lstlisting}
($\,\gg$ [p]) $\longleftrightarrow$ ($\,\tturnstile$ [p])
\end{lstlisting}
This means that the sequent calculus can prove all the valid formulas. Furthermore, any formula that is proved by the sequent calculus is valid.

We consider first the soundness and completeness of a tableau calculus whose rules are the dual of our sequent calculus rules.
We say a tableau is closed if every branch terminates in a leaf rule.
The Isabelle theory \texttt{FOL\_Tableau} contains the tableau formalization. 
The proof of \texttt{TC\_soundness} goes by induction on the inference rules for an arbitrary function denotation.
Only two of the cases cannot be proven automatically by Isabelle.
The theorem \texttt{tableau\_completeness} states the completeness property.

The Isabelle theory \texttt{FOL\_Sequent} contains the sequent formalization. 
The lemma \texttt{SC\_soundness} states the soundness property.
Dually to the tableau, there are only two cases that are not solved automatically by Isabelle.
Finally we show a correspondence between the tableau and sequent calculus which allows to prove completeness of the latter.
The proof goes by induction over the tableau rules with a bit of massaging of each induction hypothesis to make us able to apply the corresponding sequent calculus rule.
The theorem \texttt{SC\_completeness} states the completeness property.

\section{Teaching the Sequent Calculus}

We use the formalization in Isabelle/HOL of the Sequent Calculus Verifier (SeCaV) in a bachelor course on logic for computer science and in this section we discuss our experiences.

In fall 2019 we had 52 students in the DTU course 02156 Logical Systems and Logic Programming (5 ECTS). It is not a mandatory course. Each week we offer 2 hours of lectures and 2 hours of exercises, for 13 weeks in total, plus 4 mandatory assignments and a 2 hour written exam without computer. 

We cover some example proofs in the lecture and explain in details the formalization in Isabelle/HOL of the Sequent Calculus Verifier (SeCaV) available here:
\url{https://bitbucket.org/isafol/isafol/src/master/Sequent_Calculus/SeCaV.thy}

In the first week the student proves the following 7 formulas:
$$
\begin{array}{c}
p \rightarrow p
\quad\quad
p \rightarrow \neg \neg p
\quad\quad
\neg \neg p \rightarrow p
\quad\quad
\forall x p(x) \rightarrow p(a)
\quad\quad
\forall x p(x) \rightarrow \exists x p(x)
\\[1ex]
p \rightarrow q \rightarrow p
\quad\quad
p \land (p \rightarrow q) \rightarrow q
\end{array}
$$

In the second week the student proves the following 7 formulas:
$$
\begin{array}{c}
(p \rightarrow q) \rightarrow p \rightarrow q
\quad\quad
p \rightarrow (p \rightarrow q) \rightarrow q
\quad\quad
\forall x \forall y p(x, y) \rightarrow \forall x p(x, x)
\quad\quad
p \land q \rightarrow r \rightarrow p \land r
\\[1ex]
p(a) \land (p(a) \rightarrow \forall x p(x)) \rightarrow \forall x p(x)
\quad\quad
\neg p \lor \neg q \rightarrow \neg(p \land q
\quad\quad
(p \rightarrow q \rightarrow r) \rightarrow (p \rightarrow q) \rightarrow p \rightarrow r
\end{array}
$$
This year we recommended using refutation and the systematic construction of a semantic tableau before proving the formula in the sequent calculus.
It is possible to prove each of the formulas in 16 or fewer steps.
The students prove the formulas on paper with the formalization as a reference.
We display the main definitions on the projector screen and solve example proofs on the blackboard.
All in all it is a nice mixture of old and new technology and we have two teaching assistants in the room in order to help the students.

As part of the final assignment we asked two questions involving SeCaV:

\begin{itemize}
\item 
Question 1

Consider the following formula: $(\forall x p(x) \land \forall x q(x)) \rightarrow \forall x (p(x) \land q(x))$

Use refutation and the systematic construction of a semantic tableau.
State whether this shows that the formula is valid or not.
Show a proof in the Gentzen system $\cal G$ if the formula is valid --- and prove the formula in SeCaV too.

\item
Question 2

Prove the following formulas in SeCaV using the steps in the previous question:
\begin{itemize}
\medskip
\item
$(\forall x p(x) \land \forall x q(x)) \rightarrow \forall x (p(x) \land q(x))$
\medskip
\item
$p \land q \rightarrow q$
\medskip
\item
$p(a,a) \rightarrow \exists x \exists y p(x,y)$
\medskip
\item
$(\forall x p(x) \lor \forall x q(x)) \rightarrow \forall x (p(x) \lor q(x))$
\medskip
\item
$p \lor (p \rightarrow q)$
\medskip
\item
$(p \rightarrow q) \lor (q \rightarrow r)$
\medskip
\end{itemize}
\end{itemize}

We have collected anonymous answers to two questionnaires.
In the first week we had 26 answers and in the second week we had 23 answers.
Even though the course has 52 students we estimate that only 45 students are active.
However, at our university a course becomes mandatory for a student upon registration for the exam the first time.
Furthermore, at most three exam attempts are allowed.
We are aware that many of the students follow another course with a large assignment exactly in the second week of our coverage of the sequent calculus.

Most students are in the 5th semester (3rd year of the bachelor programme). A few are just in the 3rd semester and some are in the 7th semester or later.
About 60\%\ are on the BSc in Software Technology programme.
Almost none had used a proof assistant.

We asked the students about how well they understood the Sequent Calculus Verifier (SeCaV), just after a short repetition of the topics of the first week and again after further explanations in the second week.
On a scale from 0 to 9, where 0 is not-at-all, 5 is medium and 9 is fully, the median increased from 4 to 5 but still with answers from 0 to 7.
The average started at 3.6 and ended at 4.0 (note that this is still before the exercise session in the second week).
For comparison, we also asked about the understanding of a Hilbert system for propositional logic --- here the average was 5.0 (median 6 with answers from 0 to 7).

Statistics about the course evaluations and the course grades are publicly available.
The course evaluation was very positive. The questions were changed recently so it is difficult to compare with previous years.

In Denmark we use a so-called 7-step-scale, designed to be compatible with the ECTS grading scale:

\begin{enumerate}
\item[A]
For an excellent performance displaying a high level of command of all aspects of the relevant material, with no or only a few minor weaknesses.
\item[B]
For a very good performance displaying a high level of command of most aspects of the relevant material, with only minor weaknesses.
\item[C]
For a good performance displaying good command of the relevant material but also some weaknesses.
\item[D]
For a fair performance displaying some command of the relevant material but also some major weaknesses.
\item[E]
For a performance meeting only the minimum requirements for acceptance.
\item[Fx]
For a performance which does not meet the minimum requirements for acceptance.
\item[F]
For a performance which is unacceptable in all aspects.
\end{enumerate}

The grades have improved this year and we think that this could be related to the use of SeCaV.

\begin{center}
\begin{tabular}{lrrrrrrrr}
Year & A & B & C & D & E & Fx & F & Total \\
2013 & 12 & 14 & 13 & 7 & 1 & 6 & 1 & 54 \\
2014 & 11 & 13 & 21 & 8 & 0 & 8 & 0 & 61 \\
2015 & 14 & 14 & 25 & 5 & 0 & 3 & 0 & 61 \\
2016 & 18 & 17 & 11 & 16 & 0 & 2 & 0 & 64 \\
2017 & 11 & 23 & 23 & 11 & 0 & 5 & 0 & 73 \\
2018 & 10 & 16 & 13 & 1 & 0 & 5 & 1 & 46 \\
2019 & 20 & 13 & 6 & 4 & 0 & 1 & 1 & 45
\end{tabular}

\end{center}

The slight decrease in the number of students might be due to the introductory machine learning course in the same time slot. That course has increased from 88 students in 2013 to 533 students in 2019.

\section{Related Work}
A number of formalizations in proof assistants of sequent calculi appear in the literature. Ridge and Margetson \cite{Completeness-AFP,ridge,Ridge05} formalized in Isabelle a sequent calculus that is also implemented as a verified prover. The calculus is for classical first-order logic formulas in negation normal form and the language of terms consists of only variables. Blanchette, Popescu and Traytel \cite{blanchettepopescutraytel} formalized in Isabelle a general framework for soundness and completeness proofs. In the supplementary material to another paper \cite{blanchettepopescu}, Blanchette and Popescu provide an instance with a formalized tableau for many-sorted first-order logic in negation normal form with equality, but this is not kept up to date with the current version of Isabelle.
Braselmann and Koepke \cite{braselmanncalc,braselmanncomplete} formalized in Mizar a sequent calculus for classical first-order logic and proved it sound and complete. Schl\"oder and Koepke \cite{schloder} proved it complete for also uncountable languages.
Ilik, Lee and Herbelin \cite{ilikleeherbelin} introduced a Kripke-style semantics for classical first-order logic, and Ilik \cite{ilik} formalized in Coq the completeness of a sequent calculus with respect to this semantics.
Persson \cite{persson} formalized, in ALF, a sequent calculus for intuitionistic first-order logic and proved it sound.
Herbelin, Kim and Lee \cite{herbelinkimlee} formalized in Coq a sequent calculus for intuitionistic first-order logic with implication and universal quantification as the only logical symbols, and proved it sound and complete with respect to a Kripke-style semantics.

Formalizations of other proof systems for first-order logic also appear, such as axiomatic systems for classical logic (in Isabelle by Jensen, Larsen, Schlichtkrull and Villadsen \cite{jensen,spa19}), natural deduction for classical logic (in Isabelle/HOL by Berghofer \cite{berghofer,nadea19,nadea} and in Phox by Raffali \cite{raffalli}), natural deduction for intuitionistic logic (in ALF by Persson \cite{persson}), resolution (by Schlichtkrull \cite{schlichtkrull} in Isabelle/HOL and also in Isabelle/HOL by Schlichtkrull, Blanchette, Traytel and Waldmann \cite{schlichtkrullblanchettetraytelwaldmann}) and superposition (by Peltier \cite{peltier} in Isabelle/HOL).
Paulson's formalization in Isabelle/HOL of G\"odel's Incompleteness Theorems \cite{Incompleteness-AFP} does not include a proof of completeness of a proof system.

We are not aware of other educational uses of proof assistants based on formalized soundness and completeness proofs, with the exception of our Natural Deduction Assistant (NaDeA) \cite{nadea19}. We first and foremost developed the Sequent Calculus Verifier (SeCaV) in order to obtain a simpler approach to teaching logic, in particular the intuitionistic flavor of natural deduction can complicate proofs and in general sequent calculus is better geared towards automation.

\section{Conclusion and Future Work}

We have presented a formalization in Isabelle/HOL of a sequent calculus for first-order logic with a programming-like approach to the syntax, the semantics and the proof system.
The resulting system, called Sequent Calculus Verifier (SeCaV), is different from our Natural Deduction Assistant (NaDeA) \cite{nadea19}, a tool for teaching natural deduction rather than sequent calculus.

We use the formalization in a bachelor course on logic for computer science and have polished the soundness and completeness proofs in order to make them easier to understand for students and researchers.
The formalization of the sequent calculus uses no higher-order functions, that is, no function takes a function as argument or returns a function as its result. This can be advantageous in a bachelor course, in particular if the students are not familiar with functional programming.

Future work also includes the extension to a prover (a program that tries to prove a formula using the sequent calculus), perhaps by extending our verified simple prover for first-order logic without constants and functions \cite{VilladsenEtAl:PAAR2018}.

\appendix

\section*{Appendix: Formalization in Isabelle/HOL}\label{app:inductive}

We show the complete formalization of the sequent calculus except for the proofs of soundness and completeness (hence about 5000 lines are omitted including the formalizations of tableaux).

\newpage

\noindent
First the syntax and the Herbrand terms used in the completeness theorem:

\

\noindent\includegraphics[trim=10mm 198mm 38mm 10mm,clip,width=
\textwidth,height=\textheight,keepaspectratio,page=1]{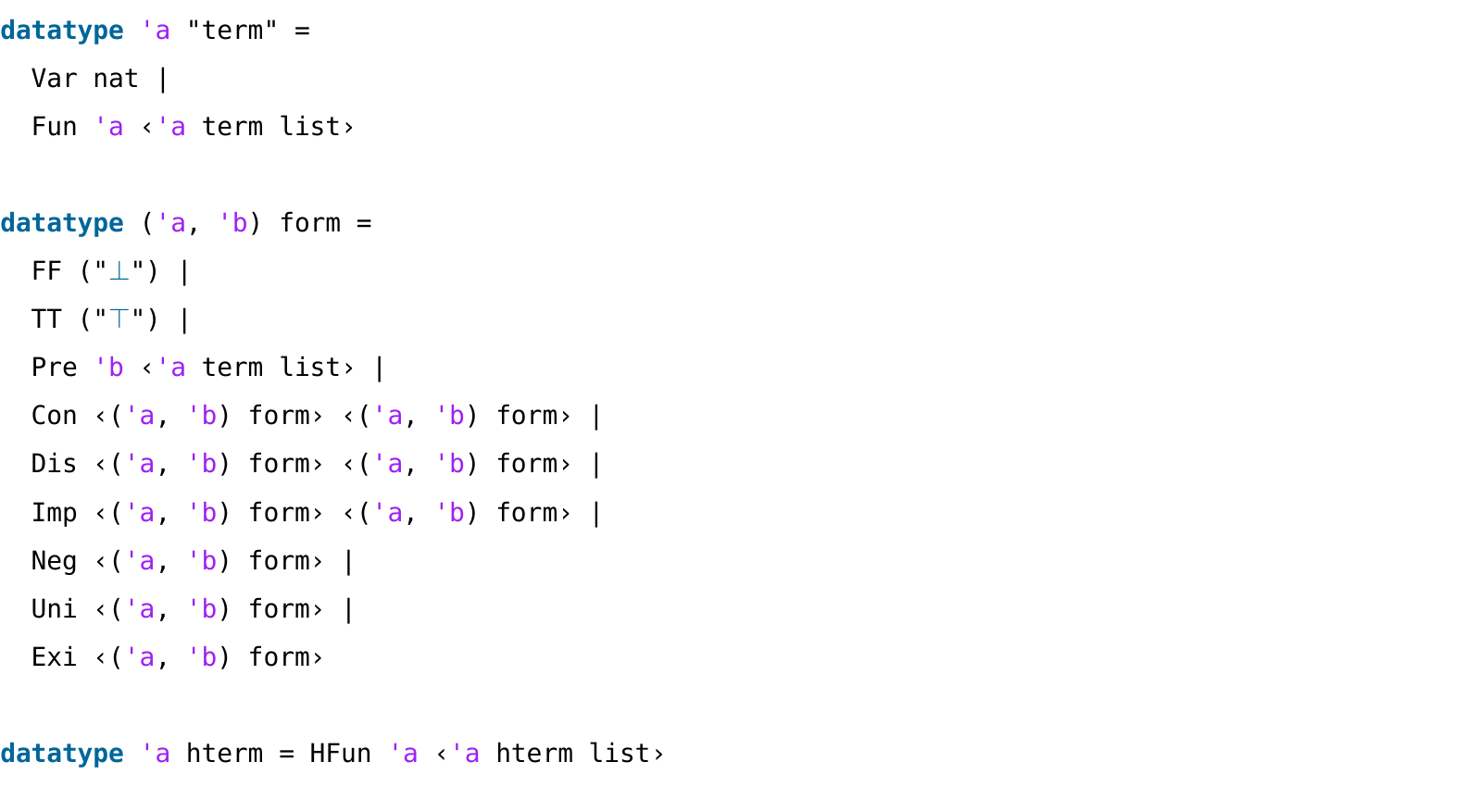}

\noindent
Then the semantics (with a definition for handling the variable environment):

\

\noindent\includegraphics[trim=10mm 182mm 38mm 10mm,clip,width=\textwidth,height=\textheight,keepaspectratio,page=2]{Appendix}

\newpage

\noindent
Auxiliary functions for new constants and functions:

\

\noindent\includegraphics[trim=10mm 176mm 38mm 10mm,clip,width=\textwidth,height=\textheight,keepaspectratio,page=3]{Appendix}

\newpage

\noindent
Auxiliary functions for substitution for variables:

\

\noindent\includegraphics[trim=10mm 166mm 38mm 10mm,clip,width=\textwidth,height=\textheight,keepaspectratio,page=4]{Appendix}

\

\noindent
Auxiliary functions for expanding a list of formulas (thinning and reordering):

\

\noindent\includegraphics[trim=10mm 244mm 38mm 10mm,clip,width=\textwidth,height=\textheight,keepaspectratio,page=5]{Appendix}

\newpage

\noindent
The sequent calculus as an inductive definition:

\

\noindent\includegraphics[trim=10mm 196mm 38mm 10mm,clip,width=\textwidth,height=\textheight,keepaspectratio,page=6]{Appendix}

\noindent
The main theorem with the proofs omitted:

\

\noindent\includegraphics[trim=10mm 260mm 38mm 10mm,clip,width=\textwidth,height=\textheight,keepaspectratio,page=7]{Appendix}

\noindent
And finally a straightforward corollary:

\

\noindent\includegraphics[trim=10mm 277mm 38mm 10mm,clip,width=\textwidth,height=\textheight,keepaspectratio,page=8]{Appendix}

\noindent
The corollary is not as general as the theorem with respect to soundness.

\

\noindent
A simple proof of the corollary is \,``\texttt{using complete\_sound by fast}''\, and it can be found automatically using the Sledgehammer panel.

\

\noindent
It is possible to check the entire formalization by opening the following file \url{https://bitbucket.org/isafol/isafol/src/master/FOL_Berghofer/FOL_Appendix.thy} in Isabelle/HOL.

\newpage

\bibliographystyle{eptcs}
\bibliography{references}

\end{document}